Site-dependent conduction channel transmission in atomic-scale superconducting junctions

Howon Kim and Yukio Hasegawa*

The Institute for Solid State Physics, the University of Tokyo

5-1-5, Kashiwa-no-ha, Kashiwa 277-8581, Japan

abstract

Using scanning tunneling microscopy, we reproducibly obtained electrical conductance spectra of superconductor- superconductor atomic point contacts formed on various atomic sites of the substrate. From the analysis of the subharmonic features due to the multiple Andreev reflection, spatial distributions of the number of the conduction channels and their transmission probabilities were obtained. It is found that the number of channels and the transmission probabilities strongly depend on an atomic site where the contact is formed. We also revealed how the conduction channels and their transmission probability evolve from the tunneling to contact regimes. The Josephson current also exhibits the atomic site dependence, which is explained by the site-dependent transmission probabilities. Our results demonstrate a crucial role of the atomic geometry in the conduction channels of the ultimately confined conductor for the first time experimentally.

When an atomic-size point contact is formed between two metal electrodes, quantum phenomena appear in its electrical conductance. The coherent transports through the junction can be described as the sum of the contributions from individual eigenmodes called conduction channels [1-3]. A set of the transmission probabilities of the channels is, thus, regarded as a *personal identification number* (PIN) code of the junction. According to the famous Landauer formula, the total normal-state conductance $G_N$ is given as $G_N = G_0 \sum_i^N \tau_i$. Here, $N$ is the number of the channels, $\tau_i$ is the transmission probability of $i$-th channel, and $G_0$ is the quantum conductance given by $2e^2/h$ ~ 77.5 μS, where $e$ is the elementary charge and $h$ is Planck's constant. In the case of atomic-scale superconductor-superconductor (S-S) junctions, the supercurrent due to the Josephson effect (a zero-bias peak (ZBP) in *dI/dV* of Fig. 1(b)) [4, 5] and the conductance peaks at fractional energies of the superconducting gap due to multiple Andreev reflection (MAR) [6-13] (peaks at $2\Delta/e$ and $\Delta/e$ and a shoulder at $2\Delta/3e$ in *dI/dV*, where $\Delta$ is the superconducting gap of the two junction-forming electrodes) can also be described with the $\{\tau_i\}$ set. In this Letter, through the conductance analyses of S-S atomic point-contacts formed reproducibly by scanning tunneling microscopy (STM), the site and gap-distance dependences of the $\{\tau_i\}$ sets were obtained in atomic scale for the first time. From the results we revealed how the conduction channels open and the transmission probabilities evolve with the gap distance from the tunneling to the contact regimes. We also found for instance that the number of the conduction channels between the contacts formed *on an atom* (on-top site in Fig. 1(a)) and *between atoms* (fcc and hcp hollow sites) are different.

Experimentally a $\{\tau_i\}$ set has been obtained from the shape of the subharmonic gap structures due to MAR, which can be observed in conductance spectra of S-S point contacts, by using mechanically controlled break junction (MCBJ) [10, 11, 14] and STM [11, 15-17]. In the previous studies, however, atomic geometry of the point contact was not well defined. Since the conductance and channel transmission should be sensitive to the atomic geometry, as pointed out theoretically [14, 18], the obtained $\{\tau_i\}$ sets varied in every measurement and were thus often analyzed statistically. In the MCBJ method, controlling the contact atomic geometry is technically impossible, and the atomic resolution of STM has not been utilized for defining the site of the point contact. A $\{\tau_i\}$ set of a junction with specified atomic geometry, therefore, has not been obtained experimentally yet.

In this study using a cryogenic STM (Unisoku, USM-1300S) with a controller (Specs, Nanonis) [19, 20] we formed S-S atomic point contacts at 1.62 K between a superconducting Pb-coated PtIr tip and (111)-oriented Pb island structures whose thickness is > 25 monolayer formed on Ge(111)-$\beta\sqrt{3}\times\sqrt{3}$ Pb phase. The crystallographic orientation of the island structures and the two hollow sites were identified from atom-manipulation images taken on the same island [19]. The tip was coated with Pb by intentional indentation into a Pb island structure. STM images taken on the sample, seen in Sec. 1 of Supplementary Materials (SM) [21], exhibit a moiré structure due to the lattice mismatch between the Pb layer and substrate. In order to avoid influence of the moiré, the conductance measurements presented in this paper were performed in low-contrasted area of the moiré. The *dI/dV* spectra were taken in a standard lock-in method with the modulation frequency of 971 Hz and amplitude of 50 $\mu V_{RMS}$.

The atomic geometry of the junction was defined precisely by positioning the tip at on-top, fcc, or hcp sites of the crystalline lattice of the substrate, as schematically shown in Fig. 1(a), on an atomically resolved STM image taken prior to the measurements (See Fig. S1(b) for atomically resolved STM image [21]). After the site positioning, $I$-$V$ and $dI/dV$ spectra were taken after every motion of the tip toward the substrate by 5 pm until the tip displacement $\Delta z$ reaches -50 pm. In this way, we obtained an evolution of the spectra with the tip distance from tunneling to contact regimes above the specific atomic sites of the substrate. Since we obtained an atomically resolved image before and after the spectral measurements, the point contact was presumably formed by a single atom of the tip apex. At the original tip height ($\Delta z = 0$) set with the substrate bias voltage $V_s$ of 6.2 mV and the tunneling current $I_t$ of 50 nA, the on-top site is higher than the hollow sites by 3-4 pm, as measured from STM images taken with the condition. The tip height measured from the surface atomic plane of the substrate is thus different by this amount between the on-top and hollow sites for the same tip displacement $\Delta z$.

The obtained $I$-$V$ and $dI/dV$ spectra taken above the three sites are shown in Figs. 2(a) and 2(b), respectively. Two peaks at ± 2.4 mV in the $dI/dV$ spectra are the superconducting coherence peaks corresponding to the sum of the superconducting gaps of the two Pb electrodes. A ZBP due to the Josephson current and the subharmonic gap structures due to MAR, *e.g.* peaks at ± 1.2 mV, are also clearly visible in the spectra. The site-dependent evolution of the normal-state conductance $G_N$, which was measured outside the gap, was consistent with that reported in our previous study [19], as shown in Fig. 2(c) (SM Sec. 2 for details [21]).

A set of transmission probabilities {$\tau_i$} of the conduction channels can be determined experimentally from an analysis of the MAR subharmonic structures [6-13]. In the Andreev reflection, an electron with energy $E$ injected into the superconducting gap is reflected as a hole with energy of $E_F$-$E$ ($E_F$ is the Fermi energy) accompanied by generation of a Cooper pair in the condensate. When the bias voltage larger than $2\Delta/e(n+1)$ is applied between the S-S junction, $n$-time Andreev reflections make a quasiparticle transfer between the electrodes with the total charge transfer of $e(n+1)$, and induce a current onset at the threshold voltage. Since the probability of the Andreev reflection of $i$-th channel is given by $\tau_i$, the total current and its spectral shape due to MAR significantly depends on the $\{\tau_i\}$ set. From a fitting of the subharmonic structures, therefore, the $\{\tau_i\}$ set can be obtained [6-13].

Using a method described in Ref. 13, we fitted the measured $I$-$V$ spectra with calculated ones, as shown in Fig. 3(a), and extracted $\{\tau_i\}$ from each spectrum as presented in Fig. 3(b). Before the fitting the ZBP due to the supercurrent was removed by using the Ivanchenko-Zil'berman (IZ) equation [22], as shown in the inset of Fig. 2(a). Details of the IZ equation will be discussed below. From the obtained $\{\tau_i\}$ sets, we found several features on the evolution of the conduction channels and its site dependence, which have not been observed before. In the tunneling regime, where the normal conductances varies exponentially with the gap displacement ($\Delta z = 0$pm ~ -20pm in Fig. 2(c)), the transport is governed by a single channel, as expected. When the conductance start to deviate from the exponential line ($\Delta z = -25$ pm), the contribution of the second channel becomes significant. At $\Delta z = -30$ pm, where the conductance at on-top site becomes larger than the other sites, that is, where the chemical interaction is exerted between the tip-apex atom and the on-top atom [19], the $\tau$ value of the first and second channels is enhanced at on-top site compared with the other sites. At -35 pm, where the all conductance traces crossover, the $\tau$ distributions of all sites seem similar including the third channel. At closer tip-sample gap distances, where the increment of the conductance is suppressed and therefore it is reasonably called contact regime, $\tau$ of all channels increases with those of hcp largest among the three sites. Curiously, the two hollow sites have a 4th channel contribution whereas on-top site does not, and among the two hollow sites hcp site has larger 4th channel contribution than fcc. It was reported that a Pb-Pb atomic contact has three conduction channels [11, 23], which is equal to the number of the valence orbitals of the element. Whereas the atomic geometry in the previous studies was not specified, our experimental results directly prove that the number of channels depends on the contact site, the gap distance, and atomic geometry. In our previous paper [19] we pointed out a significant role of the chemical interaction between the tip-apex atom and substrate atom(s) in the conductance of atomic point contact, and the conductance difference comes from different number of atoms involved. These differences probably contribute to the site dependence of the $\{\tau_i\}$ set including the 4th channel.

For the spatial variation of the $\{\tau_i\}$ set, we also performed the spectral measurements of the atomic point contact during scanning over the substrate. At every pixel of the STM imaging, we made an atomic point contact by forwarding the tip toward the substrate by 50 pm, took the spectra, and then retreat the tip back to original feedback condition ($V_s$ = 6.2 mV, $I_t$ = 50 nA). Figure 4 shows the results; an atomically resolved STM image (a), simultaneously obtained mappings of ZBP conductance (b), which is closely related with the Josephson current, and normal-state conductance $G_N$ (c), as well as color-contrasted $dI/dV$ spectra (d) and the $\{\tau_i\}$ set (e) obtained along the dotted line drawn in the topographic image (a).

The periodic variations reflecting the atomic structure observed in these figures indicate the atomic site dependence of the transport properties. The quantitative relation of $G_N$ among the three sites, hcp>fcc>on-top [19], was well reproduced in Fig. 4(c). The *dI/dV* spectra presented in Fig. 4(d) clearly show atomic site dependence, in particular in the height of ZBP. Observed spatial variation in the $\{\tau_i\}$ set is also consistent with the results of -50 pm in Fig. 3(b) including the 4th channel contribution. The site dependence of the $\{\tau_i\}$ set is not simply explained by different $G_N$, since it cannot be normalized by $G_N$. The present results prove experimentally that the variations in the $\{\tau_i\}$ set observed in the previous measurements with MCBJ and STM were indeed due to various contact sites with different atomic geometries [11, 14, 23].

The ZBP in *dI/dV* spectra, that is, the Josephson current, also has a spatial variation as shown in Fig. 4(b) and evolves with the tip-substrate distance as shown in Fig. 2(b). As will be discussed below, the supercurrent can also be determined by the transmission probabilities. The analysis of the supercurrent thus provides a crosscheck of the $\{\tau_i\}$ set obtained from the MAR analysis. The amount of the Josephson current, described as $I_c \sin\delta$, depends on the phase difference $\delta$ between the two condensates. According to the formulation by Ambegaokar and Baratoff [24], the critical current $I_c$ is proportional to the sum of $\{\tau_i\}$, that is, $G_N$, in the case of small transmission probabilities; $I_C = [(\pi\Delta/2e)\tanh(\Delta/2k_B T)]G_N$, where $T$ is the measurement temperature, and $k_B$ is the Boltzmann constant. By using STM the proportional relation has been demonstrated in the tunneling regime, where $\tau$ is small. [25-27]

In contact regime, where $\tau$ is large, $I_c$ is not proportional to $G_N$ anymore and depends on the individual values of $\tau_i$. The dependence of the supercurrent on the individual $\tau_i$ has been discussed based on the Andreev bound states [4, 5], and $I_c$ can be written as

$$I_c(\{\tau_i\}) = \sum_i^N \max_\delta \left[ \frac{e\Delta^2}{2\hbar} \frac{\tau \sin \delta}{E_A(\delta, \tau_i)} \tanh\left(\frac{E_A(\delta, \tau_i)}{2k_B T}\right) \right] \quad (1)$$

with the energy level of the bound states $E_A(\delta, \tau) = \Delta\sqrt{1 - \tau \sin^2(\delta/2)}$. The non-linear relation of $I_c$ with $\tau$, as displayed in Fig. S2 in SM [21], implies that $I_c$ depends not only on the simple summation of $\{\tau_i\}$ but also on its distribution. For a given normal-state conductance, a $\{\tau_i\}$ set that includes a large $\tau$ channel has large critical current.

In the present experimental condition, because of the electromagnetic environment coupled with the junction, the phase motion is fluctuated in a diffusive manner. For such cases, *I-V* characteristics of the supercurrent can be calculated using the approach of Ivanchenko and Zil'berman (IZ) [22] shown below;

$$I(V) = \frac{I_c^2 Z_{env}}{2} \frac{V}{V^2 + V_p^2} \quad (2)$$

, where $V_p$ is a peak energy given by $(2e/\hbar)Z_{env}k_B T_n$, considering the Johnson noise generated by an effective resistor $Z_{env}$ at temperature $T_n$. The value of $Z_{env}$ and $T_n$ was estimated as 405 Ω and 12 K from the IZ analysis of *I-V* spectra taken in the tunneling regime, as discussed in SM Sec. 3 [21]. From the fitting of *I-V* spectra taken along the dotted line in Fig. 4(a) with Eq. 2, as shown in Fig. S4(c) [21], we obtained the critical current $I_c$.

The obtained $I_c$ by the IZ model, normalized one by $G_N$ ($I_c/G_N$), and the ones calculated with the $\{\tau_i\}$ set using Eq. 1 (MAR analysis) are compared in Figs. 5(a) and 5(b). Good agreements in the plots demonstrate the consistency of the two methods and support validity of our analyses. These experimental results again prove the site dependence of the $\{\tau_i\}$ set in atomic scale, and demonstrate that the atomic geometry has a critical role on the conduction channels and their transmission probabilities in point contacts.

It has been known that the superconducting properties do not vary in a scale less than the coherence length, which is ~50 nm in the case of 25 monolyaer Pb thin films [20]. In this study we found that the critical current $I_c$ (including the one normalized by $G_N$), one of the important parameters of superconductivity, varies in atomic scale length in point contact regime. The on-top and hollow sites are separated only by 0.20 nm on the Pb(111) surface. This seems counterintuitive, but can be explained with the variation in the $\{\tau_i\}$ set. Since $\tau$ determines the ratio of the Andreev reflection, which is relevant with the phase of superconductor, to the normal one, its variation causes significant change in the superconductivity-relevant properties, such as $I_c$ and the subharmonic structures, even in a system of homogeneous $\Delta$.

As a summary, we measured conductance of S-S atomic point contacts on various atomic sites of the substrate in a reproducible manner and obtained spatial distribution of the transmission probability of the conduction channels and their evolution from the tunneling to contact regimes based on an analysis of the MAR subharmonic features. The evolution of the conduction channels was revealed for the first time, and its relation with the conductance was discussed. It is found that the channel number and the set of the transmission probabilities strongly depend on the atomic site; the hollow sites have a 4th channel contribution whereas on-top site does not, and among the two hollow sites hcp site has larger 4th channel contribution than fcc. We also found that the Josephson current also depends on the contact-forming atomic site again because of the variation of the transmission probabilities. Our measurements provide direct correlation of the point-contact conductance with its atomic geometry, which is critical for understanding the transport properties of the ultimately confined conducting structures.

Acknowledgement

The authors are grateful to Professors Gabino Rubio-Bollinger and Cristian Urbina for providing us a calculation code for the MAR analysis and Professors Mahn-Soo Choi, Takeo Kato, Drs. Noriaki Oyabu and Ryoichi Hiraoka for fruitful discussion. This work is partially funded by Grants-in-Aid for Scientific Research, Japan Society for the Promotion of Science (21360018, 25286055).

Figure captions

Fig. 1: (color online) (a) schematic of site-dependent point contact formation between a superconducting Pb-coated tip and superconducting Pb(111) substrate, (b) typical *I-V* and *dI/dV* spectra taken in superconductor-superconductor Pb atomic point contact. In the *dI/dV* spectrum, a peak due to the supercurrent is observed at zero bias voltage (zero-bias peak; ZBP). Peaks at $2\Delta/e$ and $2\Delta/2e$ and a shoulder at $2\Delta/3e$ are due to multiple Andreev reflection. Here, $\Delta$ is the amount of the superconducting gap of the both electrodes (1.18meV in this study). $G_N$ is the normal-state conductance.

Fig. 2: (color online) (a) *I-V* and (b) *dI/dV* spectra taken above on-top, fcc, and hcp sites with every 5 pm displacement toward the substrate. The differential conductance spectra were normalized by the normal-state conductance, which was measured outside the gap and written on the corresponding spectrum. The *dI/dV* spectra were offset for clarity. Inset of (a) shows a zoom of the I-V spectra around the zero bias voltage with a fitting curve by the Ivanchenko-Zil'berman equation (Eq. 2) [22]. (c) plot of the normal-state conductance $G_N$ obtained from the *dI/dV* spectra taken at on-top, fcc, and hcp sites as a function of the tip displacement $\Delta z$. Conductance traces ($G_N$-$\Delta z$ plot) taken on the three sites by measuring the current at $V_s = 6.2$ mV during the tip forward/backward motion from $\Delta z = 0$ are also presented. The inset shows zoomed plots.

Fig. 3: (color online) (a) *I-V* spectra taken above on-top, fcc, and hcp sites with various tip displacement (same as Fig. 2(b)) with fitted curves in the MAR analysis (drawn with red solid line). The supercurrent peak was removed using the Ivanchenko-Zil'berman equation (Eq. 2) [22] before the fitting, as shown in the inset of Fig. 2(a). (b) set of transmission probabilities obtained from the MAR fitting.

Fig. 4: (color online) (a) STM image ($I_t$=50 nA and $V_s$=6.2 mV) showing an atomic structure of Pb(111) surface (atomic spacing: 0.35 nm). Mappings of zero-bias peak (ZBP) conductance (b), and normal-state conductance $G_N$ (c), both of which were taken simultaneously with (a). (d) color-contrasted *dI/dV* spectra and (e) site-dependent {$\tau_i$} set taken along the dotted line drawn in the topographic image (a). The results of the MAR fitting are presented in SM Sec. 5 [21]. Major atomic sites (on-top, fcc, and hcp) are marked in the spectra and plot.

Fig. 5: (color online) (a) plot of the critical current $I_c$ as a function of the lateral tip distance obtained from the data set shown in Fig. 4 in the following two methods. One is from a fitting of the supercurrent with the IZ equation (Eq. 2) (IZ model) and the other is from the {$\tau_i$} set obtained by Eq. 1 (MAR analysis). (b) $I_c$ normalized by the normal-state conductance $G_N$ as a function of the lateral tip distance. Major atomic sites (on-top, fcc, and hcp) are marked in the plots.

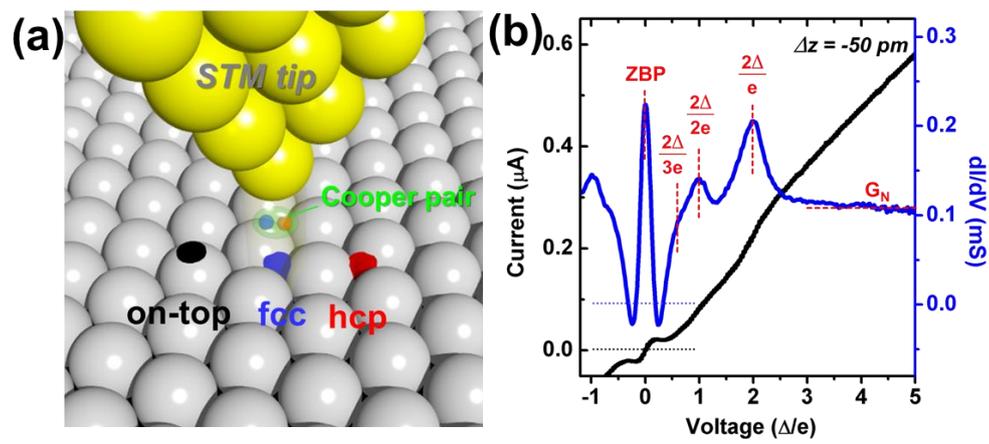

Fig. 1

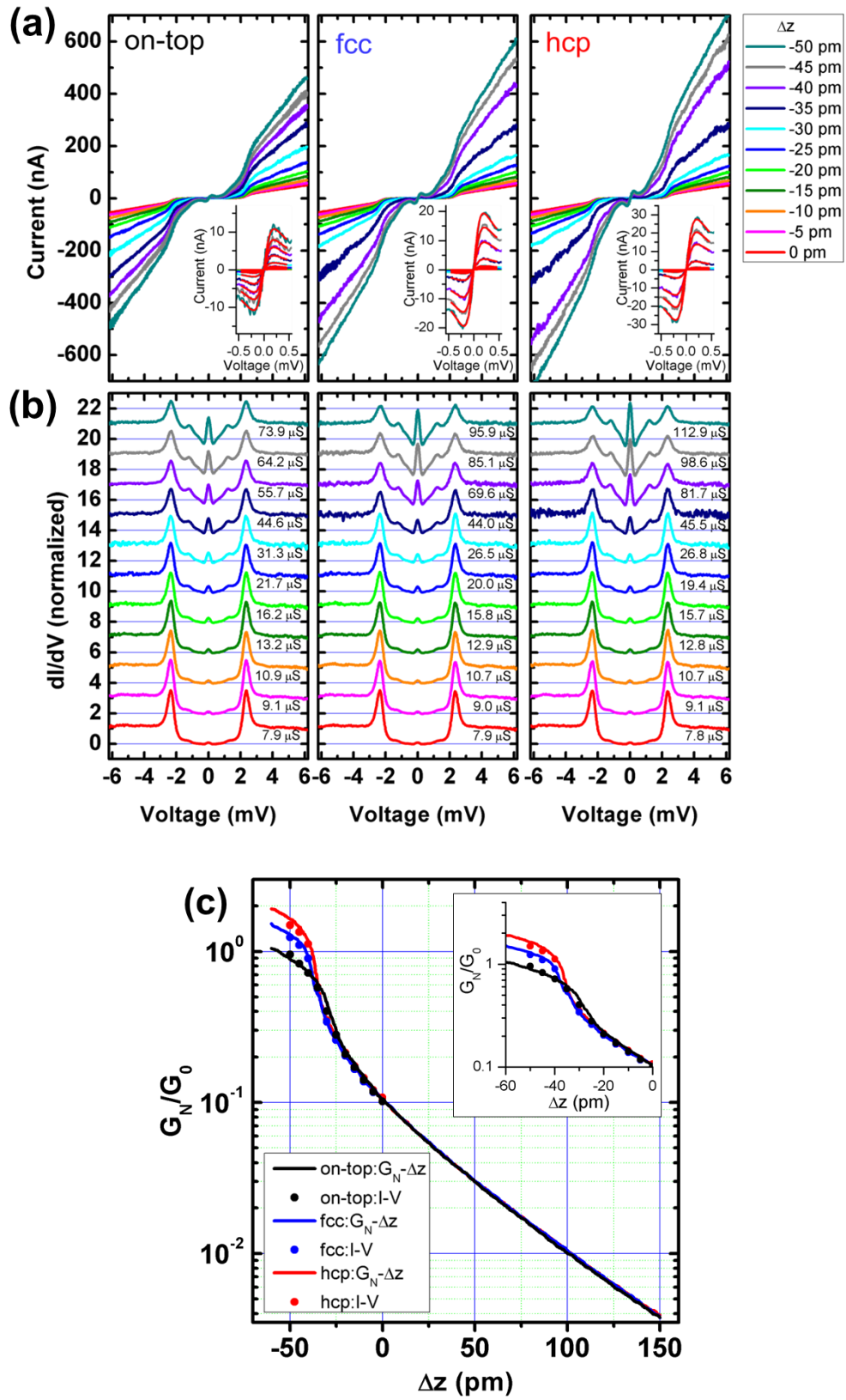

Fig. 2

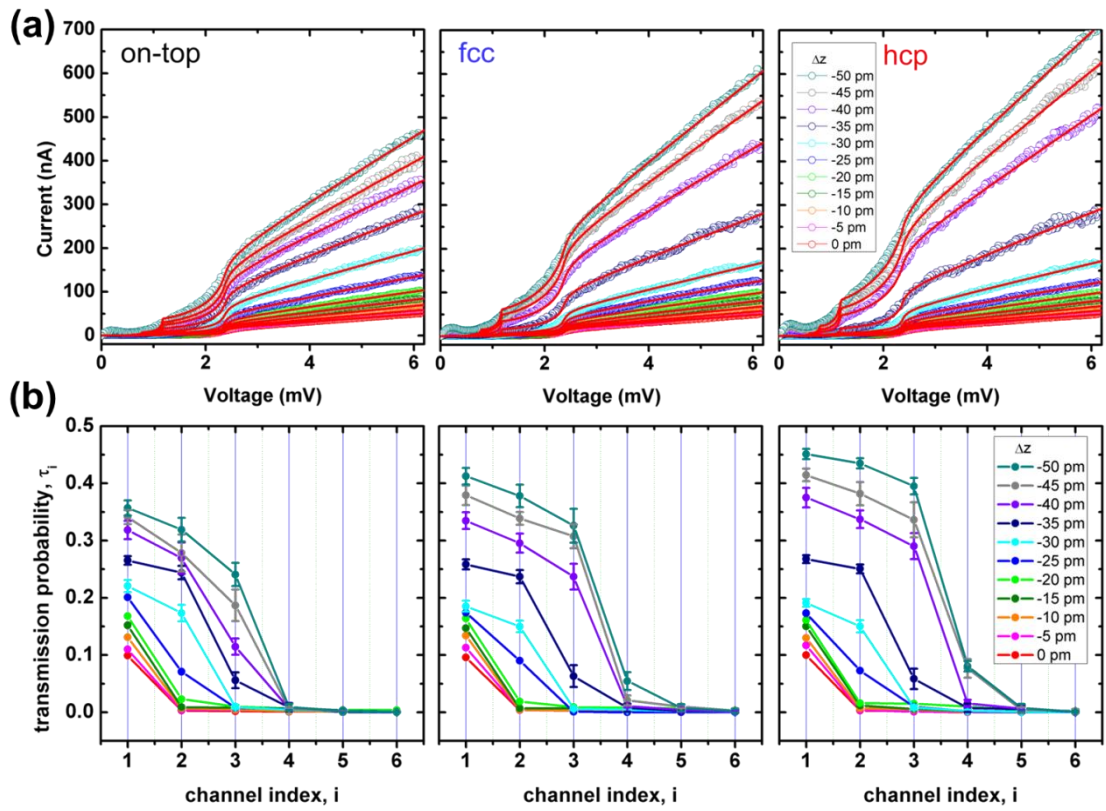

Fig. 3

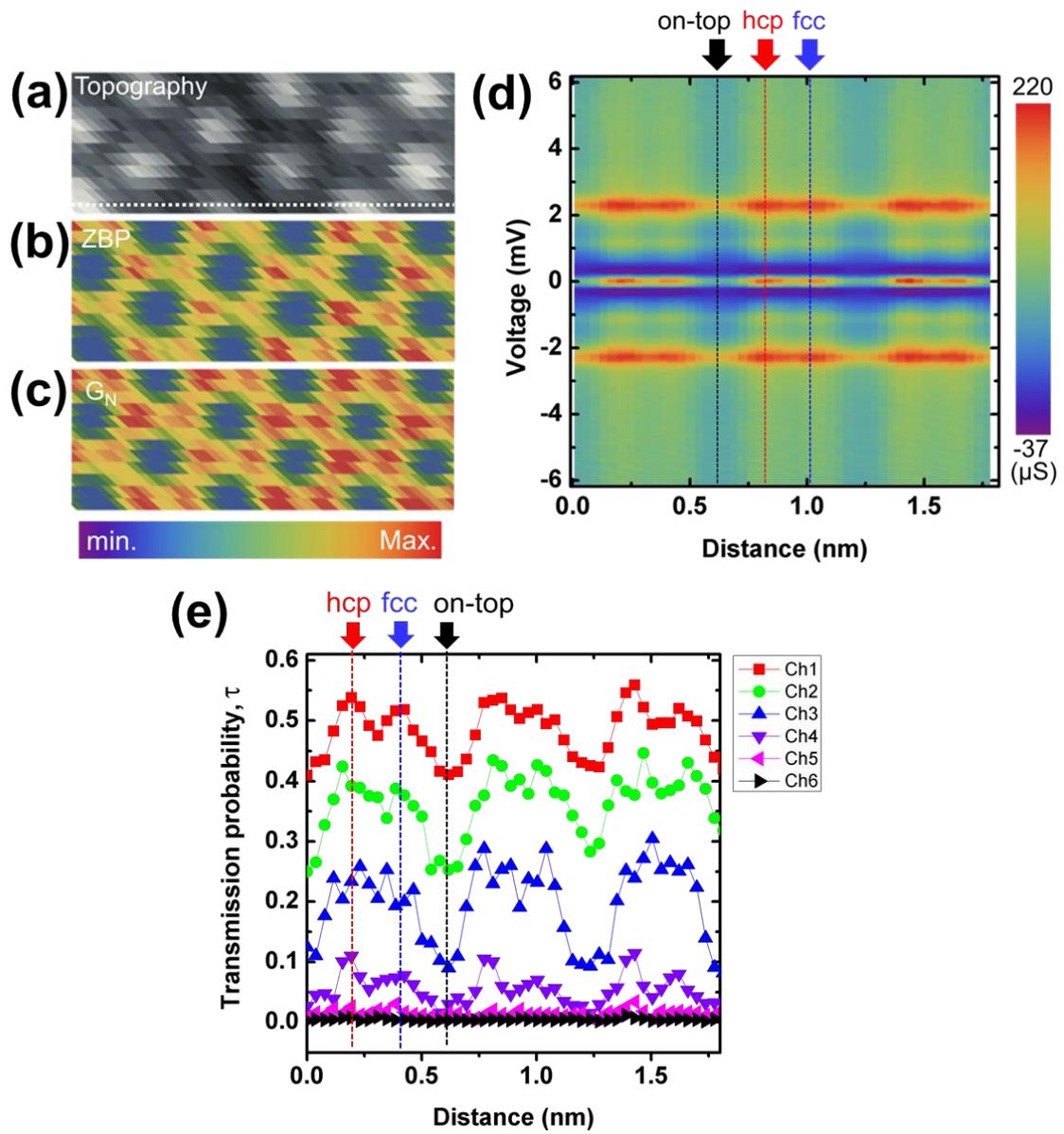

Fig. 4

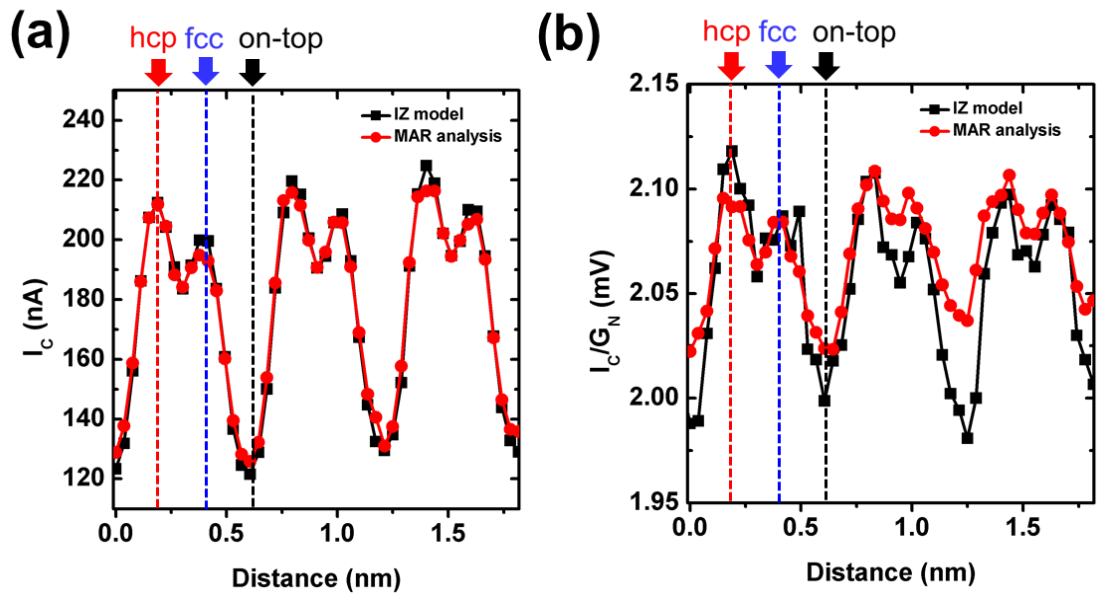

Fig. 5

Supplemental Materials of
"Site-dependent conduction channel transmission in atomic-scale superconducting junctions "
by Howon Kim and Yukio Hasegawa

**1. Sample preparation and STM images taken on the sample**

The Pb island structures we used in this study were prepared by depositing Pb on Ge(111) substrate. In order to form well-ordered island structures, we first formed Ge(111)-$\beta\sqrt{3}\times\sqrt{3}$ Pb phase on the Ge substrate and then deposited Pb to form islands. The following is a procedure of the sample preparation. All the processes were carried out in ultrahigh vacuum conditions.

1. prepare a clean Ge(111)-c(2x8) surface by Ar sputtering and subsequent annealing at 650°C
2. form Ge(111)-$\beta\sqrt{3}\times\sqrt{3}$ Pb phase by depositing ~1 monolayer (ML) Pb at room temperature and annealing the sample at ~350°C.
3. form Pb island structures by depositing ~ 3.5 ML Pb on a sample cooled at 240 K and keeping it at room temperature more than 1 hour to make the island top flat before loading the sample into a cooled STM unit.

Figure S1(a) shows typical STM images taken on a Pb island structure formed on Ge(111)-$\beta\sqrt{3}\times\sqrt{3}$ Pb phase. The island has a (111) surface on its top and the thickness is > 25 monolayer. Figure S1(b) shows an atomically resolved image taken on the (111) surface of a Pb island structure. Moiré structure due to the lattice mismatch between the Pb layer and substrate can be seen. In order to avoid influence of the moiré to the conductance, the spectra and conductance traces shown in this paper are taken in low-contrasted area of the moiré.

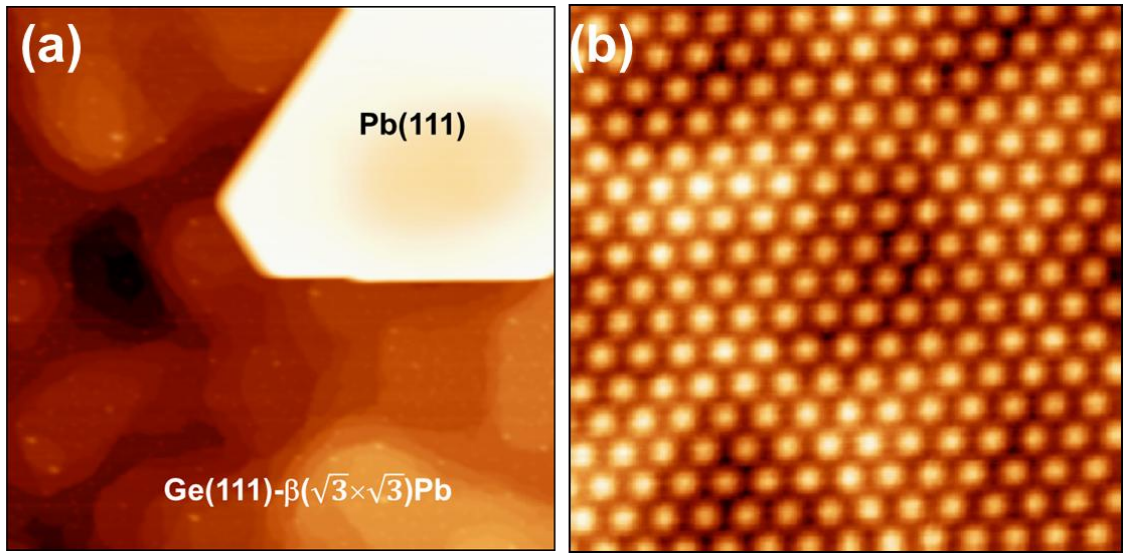

Fig. S1: (a) STM image of a Pb island structure formed on Ge(111)-$\beta\sqrt{3}\times\sqrt{3}$ Pb phase. The image was taken with the sample bias voltage $V_s$ of 50 mV and the tunneling

current of $I_t$ of 15 pA. The size of the image is 700 nm × 700 nm. (b) STM image (5 nm × 5 nm) taken on a (111) surface of a Pb island structure with $V_s$ = 9mV and $I_t$ = 50 nA.

## 2. Evolution of site-dependent normal-state conductance
Figure 2(c) shows the normal-state conductance obtained from the Fig. 2(a) data as a function of the tip-substrate displacement as well as the conductance traces ($G_N$-$\Delta z$ plot) measured at the three sites. We confirmed that the site-dependent normal-state conductance evolves in the same manner as that reported before [S1]; in the contact regime ($\Delta z$ < -40 pm) the hollow sites has larger conductance than on-top site and among the hollow sites the hcp site has larger conductance than fcc [S1]. In a transition between the contact and tunneling regimes ($\Delta z$ ~ -30 pm) the conductance at on-top site was enhanced compared with the other sites [S1] (see inset for zoom), indicating appropriate and reproducible site- and distance-dependent measurements on the Pb(111) islands, independent of the substrate materials (Ge in this study whereas Si in [S1]).

## 3. Critical current normalized by the normal state conductance
The dependence of the critical current $I_c$ on the individual $\tau$ can be written as

$$I_c(\tau) = \max_\delta \left[ \frac{e\Delta^2}{2\hbar} \frac{\tau \sin \delta}{E_A(\delta,\tau)} \tanh\left(\frac{E_A(\delta,\tau)}{2k_B T}\right) \right]$$

with the energy level of the bound states $E_A(\delta,\tau) = \Delta\sqrt{1 - \tau \sin^2(\delta/2)}$. The relation of $I_c$ normalized by the normal-state conductance $G_N$ with the transmission probability $\tau$ is shown in Fig. S2 for the case of single conduction channel.

In a limit of small $\tau$, the critical current $I_c$ is proportional to $G_N$, described as $I_C = [(\pi\Delta/2e)\tanh(\Delta/2k_B T)]G_N$, according to the formulation by Ambegaokar and Baratoff (AB) [S2]. With $\Delta = 1.18$ meV and measurement temperature $T$ = 1.62 K, the AB equation becomes $I_C = 1.85$ mV $\times G_N$. The red line corresponding to the AB relation is drawn in the plot.

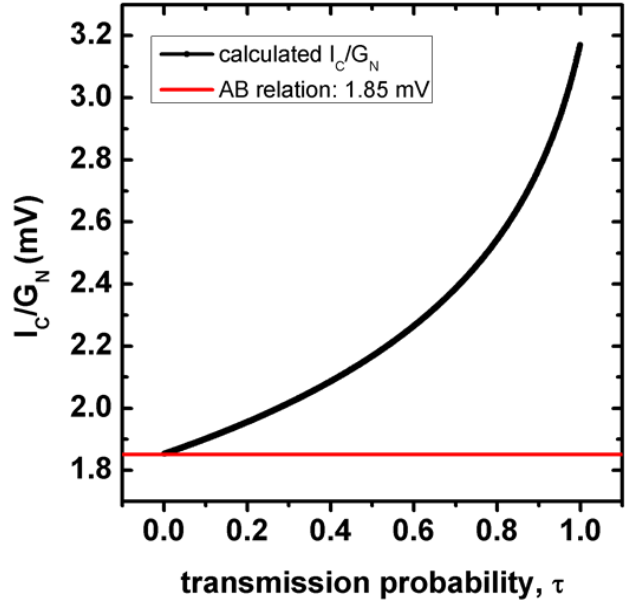

Fig. S2: critical current $I_c$ normalized by the normal-state conductance $G_N$ as a function of transmission probability $\tau$ for the case of single conduction channel. For the small $\tau$ limit, $I_c$ is proportional to the normal-state conductance $G_N$. The ratio $I_c/G_N$ deviates from the value of the Ambegaokar and Baratoff formulation (1.85 mV in this study) for large $\tau$.

### 4. $Z_{env}$ and $T_n$ used for the analysis of Ivanchenko and Zil'berman.

In order to estimate an effective resistor $Z_{env}$ at temperature $T_n$ coupled with the junction, we measured several *I-V* and *dI/dV* spectra in the tunneling regime by changing the set current of 1 nA to 800 nA as shown in the Fig. S3(a). The bias voltage applied on the substrate was set at 6.2 mV.

According to Ivanchenko and Zil'berman [S3], the supercurrent, which was observed around the zero-bias voltage, can be described as

$$I(V) = A \frac{V}{V^2 + V_p^2} \qquad A = \frac{I_c^2 Z_{env}}{2} \qquad (S1)$$

, where $V_p$ is a peak energy given by $(2e/\hbar)Z_{env}k_B T_n$, considering the Johnson noise generated by an effective resistor $Z_{env}$ at temperature $T_n$. According to the Ambegaokar-Baratoff relation [S2], which is applicable in a limit of small $\tau$, and thus in tunneling regime, $I_c$ is given as

$$I_C = \left[\frac{\pi\Delta}{2e}\tanh\left(\frac{\Delta}{2k_B T}\right)\right]G_N$$

From these equations it is found that $\sqrt{A/V_p}$ is proportional to $G_N$ in a limit of small $\tau$.

$$\sqrt{\frac{A}{V_p}} = \left[\sqrt{\frac{\hbar}{4e} \cdot \frac{1}{k_B T_n}}\frac{\pi\Delta}{2e}\tanh\left(\frac{\Delta}{2k_B T}\right)\right]G_N$$

The value of $\sqrt{A/V_p}$ was obtained from a fitting of the *I-V* spectra with Eq. S1, and plotted as a function of $G_N$. In Fig. S3(c), the $\sqrt{A/V_p}$ value taken from the I-V spectra with the set current of 10 nA to 600 nA are plotted. From the linear fitting of low $G_N$ data (set current from 10 nA to 80 nA), shown with the red line in Fig. S3(c), $Z_{env}$ and $T_n$ were estimated as 405 Ω and 12 K, respectively. We used these values for obtaining the value of $I_c$ using Eq. 2 for general $\tau$.

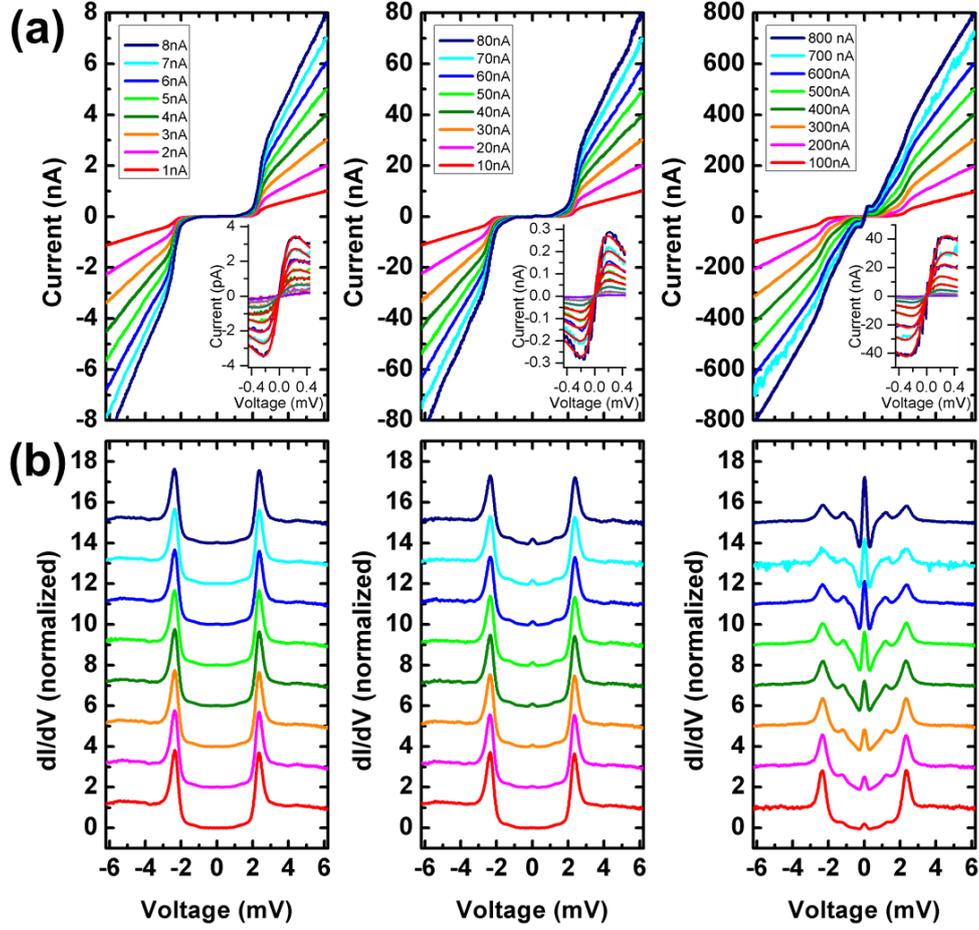

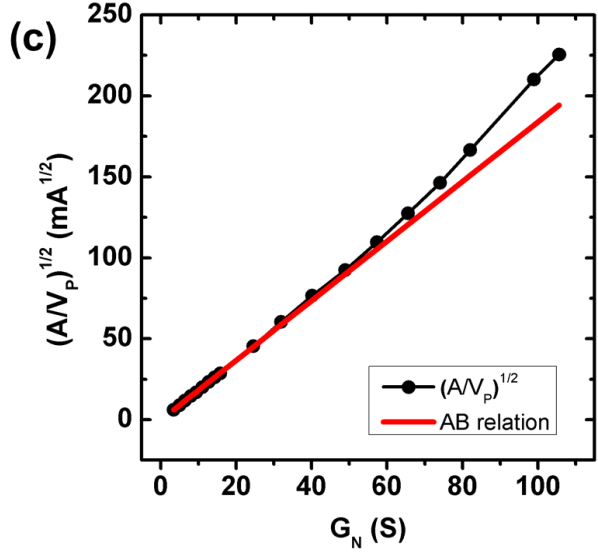

Fig. S3: (a) *I-V* and (b) *dI/dV* spectra in the tunneling regime with the set current of 1 nA to 800 nA. The bias voltage applied on the substrate was set at 6.2 mV. Inset of (a) shows a fitting of the spectra using the equation of Ivanchenko and Zil'berman (Eq. S1). (c) $\sqrt{A/V_p}$ value taken from the *I-V* spectra with the set current of 10 nA to 600 nA are plotted. From the linear fitting of low $G_N$ data (set current from 10 nA to 80 nA), shown with the red line in Fig. S3(c), $Z_{env}$ and $T_n$ were estimated as 405 Ω and 12 K, respectively.

## 5. fitting of *I-V* spectra for the analysis of multiple Andreev reflection

In order to obtain a {$\tau_i$} set, we fitted an experimentally obtained *I-V* spectrum, which shows the subharmonic gap structures, with calculated one. Before the fitting, we removed the supercurrent peak/dip around zero bias voltage using the Ivanchenko-Zil'berman equation [S3] (Eq. 2 in the main text) since the supercurrent is not relevant with the subharmonic gap structures.

*I-V* spectra shown in Fig. 3(a) are as-measured, that is, the spectra have the supercurrent structures. The supercurrent structure was removed before the MAR analysis using the IZ equation, as shown in the inset of Fig. 2(a).

Figure S4 shows the *I-V* spectra and their fitting results for the site-dependent {$\tau_i$} set displayed in Fig. 4(e), which is also shown here as Fig. S4(a). For the results, 47 spectra were fitted as shown in Fig. S4(b). In the spectra shown in Fig. S4(b), the supercurrent structure was already removed by the Ivanchenko-Zil'berman equation, as shown in Fig. S4(c).

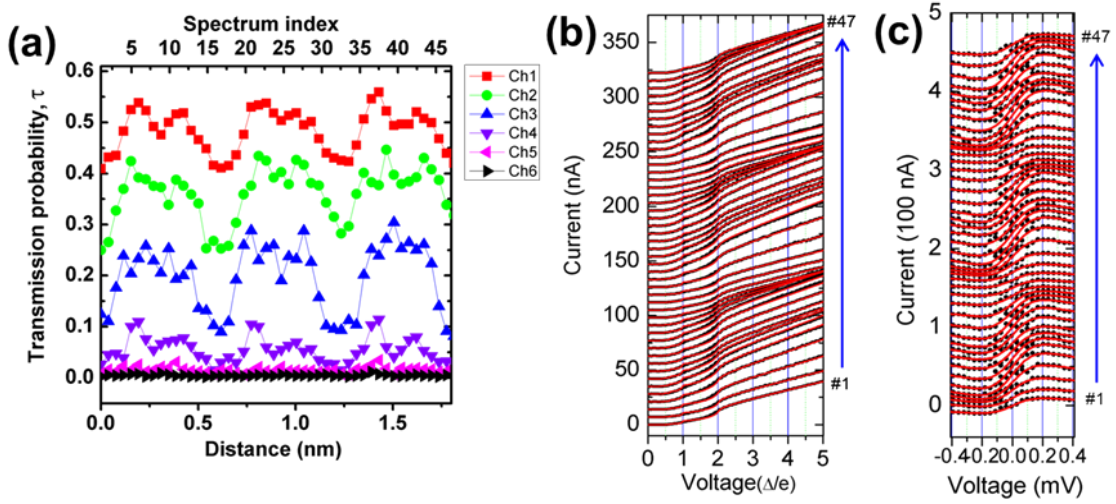

Fig. S4: (a) site-dependent transmission probabilities, same as Fig. 4(e). (b) *I-V* spectra taken along the dotted line drawn in an STM image of Fig. 4(a) with fitted curves for the MAR analysis (drawn with red line). In the spectra the supercurrent peak/dip was already removed by using the Ivanchenko-Zil'berman equation (Eq. 2 in the main text) for the fitting. From the fitting, the transmission probabilities $\{\tau_i\}$ shown in (a) were obtained. (c) *I-V* spectra around the zero-bias voltage taken along the dotted line drawn in an STM image of Fig. 4(a) with fitted curves of the Ivanchenko-Zil'berman equation (Eq. 2 in the main text).